\documentclass[pra,aps,groupedaddress,superscriptaddress,twocolumn]{revtex4-2}
\usepackage{xcolor}
\usepackage{graphicx,amsmath}
\usepackage{amssymb}
\usepackage{mathrsfs}

\begin{document}

\title{Angle and angular momentum - new twist for an old pair}

\author{Ladislav Mi\v{s}ta, Jr.}
\email{mista@optics.upol.cz}
\affiliation{Department of Optics, Palack\' y University, 17.
listopadu 12,  779~00 Olomouc, Czech Republic}
\author{Hubert de Guise}
\affiliation{Department of Physics, Lakehead University, Thunder Bay, Ontario P7B 5E1, Canada}
\author{Jaroslav \v{R}eh\'{a}\v{c}ek}
\affiliation{Department of Optics, Palack\' y University, 17.
listopadu 12,  779~00 Olomouc, Czech Republic}
\author{Zden\v{e}k Hradil}
\affiliation{Department of Optics, Palack\' y University, 17.
listopadu 12,  779~00 Olomouc, Czech Republic}

\begin{abstract}
Reaching  ultimate performance of quantum technologies requires the use of detection at quantum limits and access to all resources of the underlying physical system. We establish a full quantum analogy between the pair of angular momentum and exponential angular variable, and the structure of canonically conjugate position and momentum.  This includes the notion of optimal simultaneous measurement of the angular momentum and angular variable, the identification of Einstein-Podolsky-Rosen-like variables and states, and finally a phase-space representation of quantum states. Our construction is based on close interconnection of the three concepts and may serve as a template for the treatment of other observables.   This theory  also provides a new testbed for implementation of quantum technologies combining discrete and continuous quantum variables.
 \end{abstract}
\maketitle

{\it Introduction.}--Quantum limitations establish challenging problems for contemporary science, and  rapid progress in metrology and communications - two important pillars of our technological
world - bring us closer to this  fully unexplored ultimate regime. Though quantum effects are fundamentally distinct from our classical intuition, they are  manifested in variables which have a classical interpretation.   Conservation laws and the concept of complementary variables offer the opportunity to be safely guided   through this  unfamiliar world of intertwined quantum effects. Thus we see quantum limits more as a sophisticated network of the interconnected rules and subtle conditions rather than strict  and impenetrable barriers.

Canonical pairs of variables  like energy and time,  position and momentum, and angular momentum and angle provide the textbook examples.  For instance,
the Schr\"odinger  equation connecting the Hamiltonian with time evolution is a starting point of quantum mechanics,  whereas detection of energy of electromagnetic field at the level of single photons opened the era of quantum optics. Though these concepts are well understood, time is not an operator but a parameter controlling  the interaction, so care must be employed in understanding the energy-time uncertainty relation. The celebrated pair of position and momentum is the most famous example of  non-commuting variables and the starting point of quantum information science. The Heisenberg uncertainty principle, Einstein-Podolsky-Rosen (EPR) states \cite{Einstein_35} and their detection, coherent states and phase space representation formulated by Roy Glauber \cite{Glauber_63}, the Arthurs-Kelly concept of approximate simultaneous detection \cite{Arthurs_Kelly} (see also \cite{Stenholm}), as well as teleportation with continuous variables \cite{Braunstein_98}, are the important milestones on the  long way towards harnessing quantum effects.

The angular momentum and angular variable have been treated  similarly to the energy and time rather than  full bodied  quantum (quadrature-like) variables forming the phase space for complete description.
The purpose of this Letter is to formulate full quantum description for this conjugated pair. We show the prominent role of the minimum uncertainty states for angular momentum  and angular variable in four tasks: the formulation of saturable uncertainty relations, the simultaneous detection of non-commuting variables, the construction of EPR-like variables and states, and finally the phase-space representation of quantum states.

Our work is motivated by possible applications to metrology but more generally by overarching questions about optimal
measurements limited by the uncertainty relations. The group E(2), the natural algebraic structure for angle and angular momentum,  is an interesting testbed for the extension of techniques developed in the context of Heisenberg algebra. We mention for completeness some expressions valid for the general case of quasi-periodic representations \cite{Kastrup2006,Isham} but leave the consequences of quasi-periodicity and its potential applications (as discussed for instance in~\cite{Martin}) for later work. As  there is an extensive  body of work related to optical angular momentum as a tool for quantum information processing \cite{Molina_Terizza_07,Yao_11,Krenn_17}, the theory developed here provides theoretical framework for a full quantum description based on the concept of complementary variables as a possible new platform fully implemented on  the E(2) symmetry. Astonishing experimental progress with  sources based on structured light with imprinted optical angular momentum   \cite{Miao_16,NaturePhys21}  is a promise for the realization of such protocols  and may trigger new experimental techniques oriented to state engineering and detection at quantum limits.

{\it Universal uncertainty relations.}--Non-commutativity is an essential differentiating concept between quantum and classical
physics. We analyze  in detail the concept for the paradigmatic pair of
angular momentum $L=-i\partial_{\phi}$ and unitary exponential operator $E=e^{-i\phi}$ satisfying the commutation rule
of Euclidean algebra $\mathfrak{e}$(2): $[E,L]= E$. Rephrased in terms of Hermitian operators as ${[}S_{\alpha},L{]} = i C_{\alpha}$,
where  $C_{\alpha} = ( e^ {-i \alpha} E^\dag +e^{i \alpha} E  )/2$ and $S_{\alpha} = (e^ {-i \alpha} E^\dag - e^{i \alpha} E)/2i$, the rule implies the uncertainty relations
\begin{equation}\label{UR}
\langle(\Delta L)^2\rangle\langle(\Delta S_{\alpha})^2\rangle\ge \frac14 |\langle C_{\alpha}\rangle |^2.
\end{equation}
The corresponding minimum uncertainty states (MUS) (in the  $L$-representation) \cite{Kastrup2006,tighter}
\begin{eqnarray}\label{vM}
|n+\delta,\alpha\rangle=\frac{1}{\sqrt{I_0( 2 \kappa) }}\sum_{l\in \mathbb{Z}} e^{i (n -l) \alpha} I_{n-l} (\kappa)|l+\delta\rangle,
\end{eqnarray}
with $L|l+ \delta\rangle=(l+\delta)|l+\delta\rangle$, yield the von Mises distribution for the angle
$\phi$: $|\langle\phi|n+\delta,\alpha\rangle|^2=\exp{[2\kappa \cos(\phi-\alpha)]}/2\pi I_0( 2 \kappa).$
As a result the states $|n+\delta,\alpha\rangle$ will be  referred to as von Mises states.

Here $n+\delta$, where
$n\in\mathbb{Z}$ and $\delta\in[0,1)$, is the angular momentum mean, $\alpha$ is an  angle,
$\kappa\geq0$ represents the spread of angular variable, and $I_{n}(z)$ is the modified Bessel function \cite{Watson_44}
(see Supplemental Material Sec.~I for its definition and other properties). Note that we allow for angular momenta with
generally fractional eigenvalues $l+\delta$, whence the angular momentum eigenstates $\{|l+\delta\rangle\}_{l\in\mathbb{Z}}$ possess
quasi-periodic wave functions $\langle\phi|l+\delta\rangle=\exp{[i(l+\delta)\phi]}/\sqrt{2\pi}$ \cite{Kastrup2006}.

For fixed $\alpha, $ the  von Mises states $|n+\delta,\beta\rangle$ with $\beta\ne\alpha+k\pi$, $k\in\mathbb{Z}$, do not saturate the uncertainty relations (\ref{UR}).
However, by setting $\alpha = -\mbox{arg}\langle E\rangle$ and $ \Delta S  =  S_{  \alpha =  -\mbox{arg}\langle E\rangle}$, we get the parameter-free
uncertainty relations
 \begin{eqnarray}\label{uncertainty}
\langle(\Delta L)^2\rangle\,\omega^2    \geq\frac{1}{4} , \quad
 \omega^2  =\frac{\langle( \Delta S)^2\rangle}{|\langle E\rangle|^2 },
\end{eqnarray}
which is  saturated by {\it all} von Mises states. Importantly, the measure of the angular uncertainty
$\omega^{2}$ is complementary to angular momentum  in the sense that
\begin{equation}\label{vMmoments}
\langle(\Delta L)^2\rangle=\frac{\kappa}{2}\frac{I_{1}(2\kappa)}{I_{0}(2\kappa)},\quad \omega^2 =\frac{1}{2\kappa}\frac{I_{0}(2\kappa)}{I_{1}(2\kappa)},
\end{equation}
where $\langle E^{l}\rangle=\exp(-il\alpha)I_{l}(2\kappa)/I_{0}(2\kappa)$ derived in the Supplemental Material Sec.~II has been used.
Saturable uncertainty relations (\ref{uncertainty}) for  the complementary observables  of angular momentum   and angular variable  represent
the first important result of this Letter.

\begin{figure}
\includegraphics[width=0.67\columnwidth]{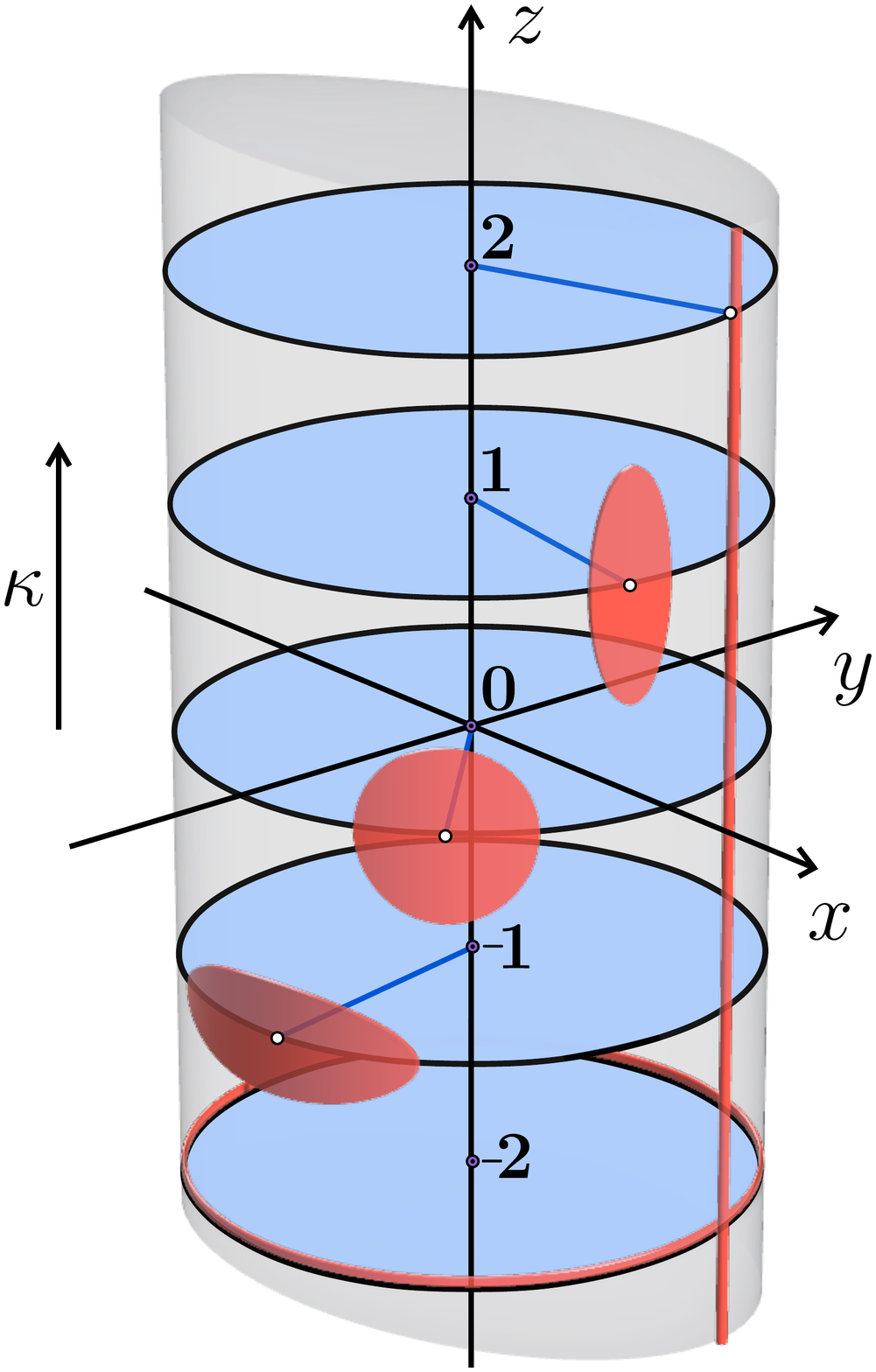}
\caption{Phase-space-representation of von Mises states $|n,\alpha\rangle$, $n\in\mathbb{Z}$, Eq.~(\ref{vM}). The pase space consists of parallel equidistant rings (black rings),
which are orthogonal to $z$-axis and their centres possess the $z$-th coordinate $n$. The von Mises state $|n,\alpha\rangle$ is represented by a noise ellipsis
(red ellipsis) centered around a point on the circle with $z$-th coordinate $n$ and polar angle $\alpha$ (positive angle between blue line segment and positive $x$-axis). The shape of
the ellipsis depends on the value of the spread parameter $\kappa$, which is chosen to grow from the bottom to the top. Accordingly, the uncertainties $\langle(\Delta L)^2\rangle$ ($\omega^2$),
Eq.~(\ref{vMmoments}), grow (decrease) from the bottom to the top. The red ring represents von Mises state with $n=-2$ and $\kappa=0$, which is an angular momentum eigenstate,
so the other phase-space rings are images of the respective angular momentum eigenstates. The red vertical line represents the von Mises state with $n=2$ in the limit of
$\kappa \rightarrow \infty$. The red circle represents the von Mises state with $n=0$ and symmetrical uncertainties $\langle(\Delta L)^2\rangle=\omega^2=1/2$ for $\kappa\doteq 1.292$.}
\label{fig1}
\end{figure}

The spread  parameter $\kappa$  has similar meaning  as ``squeezing'' but here  for the  angular momentum and  the angular variable. Since the phase space of the
pair angle and angular momentum has cylindrical topology  \cite{Kastrup2006}, one can represent von Mises states by ellipses
on the cylinder (see Fig.~\ref{fig1}) similarly to the representation of  squeezed states of a harmonic oscillator by ellipses in the plane.
Moreover, MUS  of Eq. (\ref{vM}) form an over-complete basis resolving the identity as \cite{Kastrup2006}
\begin{equation}\label{completeness}
\sum_{n\in \mathbb{Z}}\int_{-\pi}^{\pi}\frac{d\alpha}{2\pi}| n+\delta, \alpha \rangle   \langle  n+\delta, \alpha|=\openone,
\end{equation}
and can be used as a generalised measurement for the  discrete spectrum   $n+\delta$  of angular momentum  and the
continuous values $\alpha$ of the angle. In the following we set to the choice $\delta =0.$

{\it Optimal simultaneous measurement.}--The deep analogy  with $x $ and $p$ is obvious from the operator formalism behind the measurement on a signal ($s$) and ancilla ($a$) fields.
Let us define the total sum angular momentum operator and the exponential angular difference operator,

\begin{equation}  \mathcal{L}= L_s+L_a,  \quad \mathcal{E}=E_sE_a^{\dagger}.
\end{equation}
Since $[\mathcal{L},\mathcal{E}] = 0,$ the  operator $\mathcal{L}$ and any function  of  $\mathcal{E}$ and $\mathcal{E}^{\dag}$ can be  measured simultaneously and  may
serve as meter variables,  in analogy with  the pair of the EPR operators. We observe interestingly that if one assumes the unitary operator $E$ is the exponential of some ``hermitian angle'' operator, one would seemingly recover the same structure as EPR pair for  quadrature operators. However, such a conclusion  cannot be  justified here  due to the issues of periodicity.

We now move to finding optimal simultaneous measurement of  the non-commuting canonical pair $L_{s}$ and $S_{s}$. We implement the measurement via joint measurement of the commuting bipartite observables $\mathcal{L}$ and $\mathcal{S}=(\mathcal{E}^\dag -\mathcal{E})/2i$. We seek the measurement minimizing the uncertainty product $\langle(\Delta\mathcal{L})^{2}\rangle\langle(\Delta\mathcal{S})^{2}\rangle$ with $\Delta\mathcal{S}=\mathcal{S}_{\beta=\mbox{arg}\langle E_{a}\rangle-\mbox{arg}\langle E_{s}\rangle}$, where $\mathcal{S}_{\beta}=(e^{-i\beta}\mathcal{E}^\dag - e^{i\beta}\mathcal{E})/2i$, with respect to the product state $|\varphi\rangle_s|\chi\rangle_a$. A straightforward derivation with ``unbiased'' conditions $\langle  L_a \rangle   = 0, \arg \langle E_a \rangle  = 0$ and $\arg \langle E_a^{2} \rangle  = 0$
is given in the Supplemental Material Sec.~III and yields the inequality
\begin{eqnarray}\label{result2}
\langle(\Delta\mathcal{L}) ^2\rangle\langle(\Delta\mathcal{S})^2\rangle\geq\frac{1}{4}\left(|\langle E_a\rangle|+|\langle E_s\rangle |\sqrt{|\langle E_{a}^2\rangle|}\right)^2,
\end{eqnarray}
which is the second main result of this Letter. The right-hand side of the inequality represents the achievable lower bound
for the simultaneous measurement. Indeed, the inequality is saturated by the MUS  for both the system and ancilla fields satisfying the cross-condition
$\langle(\Delta L_s)^2\rangle\langle(\Delta S_a)^2\rangle=|\langle E_{a}^2\rangle|\langle(\Delta L_a)^2\rangle\langle(\Delta S_s)^2\rangle$. Consequently, the lower bound is saturated by
von Mises states $|\varphi\rangle_s=|n,\alpha, \kappa_s\rangle_{s}$ and $|\chi\rangle_a=|0,0, \kappa_a\rangle_{a}$,
with {\it different} spread parameters $\kappa_{s}$ and $\kappa_{a}$ connected by the condition
\begin{eqnarray}\label{kappacondition}
\kappa_s=\sqrt{|\langle E_{a}^2\rangle|}\kappa_a=\sqrt{\frac{I_{2}(2\kappa_{a})}{I_{0}(2\kappa_{a})}}\kappa_{a}.
\end{eqnarray}

\begin{figure}[ht]
  \includegraphics[width=0.92\columnwidth]{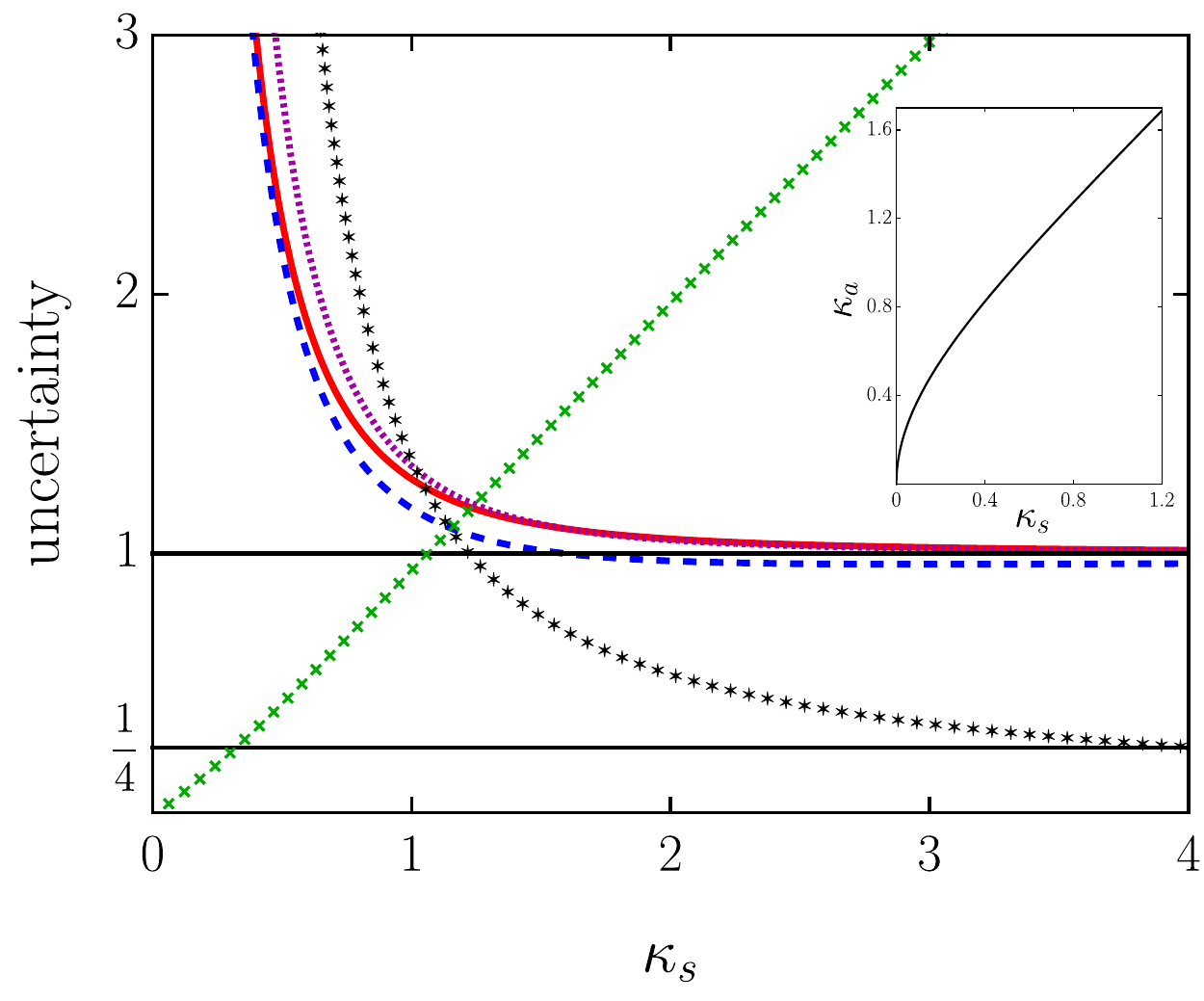}
  \caption{Uncertainties and uncertainty products for angular momentum and angular variable for optimal states and measurements
  versus the signal-state spread parameter $\kappa_s$. Uncertainties $\langle(\Delta \mathcal{L})^2\rangle$ (green crosses) and
  $\Omega^2=\langle(\Delta\mathcal{S})^2\rangle/|\langle E_{s}\rangle|^2|\langle E_{a}\rangle|^2$ (black stars), and
  uncertainty product $\langle(\Delta\mathcal{L})^2\rangle\Omega^{2}$ (solid red line) for optimal simultaneous measurement with
  von Mises signal and ancilla states with {\it different} spread parameters satisfying the optimal matching
  condition (\ref{kappacondition}) whose inverse is depicted in the inset. The same uncertainty product for suboptimal simultaneous measurement
  with von Mises signal and ancilla states with {\it the same} spread parameters $\kappa_s=\kappa_a$ (dotted magenta line).
  The product mean $\langle(\Delta\mathcal{L})^2(\Delta\mathcal{S})^2\rangle/|\langle E_{s}\rangle|^2|\langle E_{a}\rangle|^2$
  for von Mises signal and ancilla states satisfying optimal matching condition (\ref{kappacondition}) (dashed blue line).
  The uncertainty product for optimal simultaneous measurement approaches asymptotically its lower bound of $1$, which is four
  times larger than the lower bound of 1/4 for uncertainty relations (\ref{uncertainty}). The product mean
  always lies below the uncertainty product for optimal measurement and it may even lie below $1$. Equality
  $\langle(\Delta\mathcal{L})^2\rangle=\Omega^2\doteq 1.099$ is achieved for $\kappa_{s}\doteq 1.146$ and $\kappa_{a}\doteq 1.632$.}
  \label{fig2}
\end{figure}

In Fig.~\ref{fig2} we plot the optimally measurable uncertainty product (\ref{result2}) in comparison with the uncertainty relations (\ref{uncertainty}) which give the constant lower bound of $1/4$.
Note that the bound for (\ref{result2}) is  approximately 4 times larger as expected on the basis of the  Arthurs-Kelly uncertainty relations \cite{Arthurs_Kelly}, but
only in the regime where the measurement resolves the angular variable  well. This result is compared with the variance product mean $\langle(\Delta\mathcal{L})^2(\Delta\mathcal{S})^2\rangle$
derived based on the She-Heffner formalism  \cite{She_Heffner} in the Supplemental Material Sec.~V. The analysis of the latter moment normalised with respect to
$|\langle E_{s}\rangle|^2|\langle E_{a}\rangle|^2$ is surprising: it is even below the minimum value of uncorrelated product due to the anti-correlations (see dashed blue line in Fig.~\ref{fig2}).
In the words, quantum mechanics allows to specify the states (and the measurement), where each  canonically conjugated variable reaches its minimum in the uncertainty product,
but the correlated product is even below that. This indicates stronger correlations linked to the 4th order moments. Note that such an effect, though mild in our system,
is not possible in the case of $x$ and $p$ operators.

The joint measurement is realized by a projection onto
orthogonal common eigenvectors of operators ${\cal L}$ and  ${\cal E}$  corresponding to eigenvalues  $N, \Phi$ respectively. The common  eigenstates  are given as
\begin{equation}\label{EPR}
|N,\Phi\rangle_{sa}=\frac{1}{\sqrt{2\pi}}\sum_{l\in\mathbb{Z}}e^{-i l \Phi}|l+N\rangle _s|-l\rangle_a
\end{equation}
resembling the EPR-states for the position and momentum operators \cite{Einstein_35}: when  the ancilla of the state is projected onto the von Mises ancilla state
$|0,0\rangle_{a}$ with the spread parameter $\kappa$, the signal collapses into the von Mises system state $|N,\Phi\rangle_{s}$
with the same $\kappa$ as the ancilla. Below we further  develop the analogy with EPR states  by showing that the projective measurement onto the
EPR-like states (\ref{EPR}) plays the role analogous to Bell measurement for position and momentum \cite{Braunstein_98}. Generalization of the states of
Eq.~(\ref{EPR}) to signal and ancilla with generally different fractional angular momenta is discussed in the Supplemental Material Sec.~IV.
The full analogy between the structure of EPR pair and states for quadrature operators and angular momentum and  angular variable represents the third main result of this Letter.

{\it Phase-space representation.}--Existing attempts to construct a phase-space representation of angular momentum and angular variable focused exclusively on the Wigner
function \cite{Wigner_32} using group-theoretical methods \cite{Nieto_98,Kastrup2016} or employing analogies with the harmonic
oscillator \cite{Plebanski_00,Rigas_08}. Building on the latter ideas and our previous results, we develop a
complete hierarchy of phase-space distributions exhibiting behaviours and connections very much like the quasiprobability distributions
of the standard harmonic oscillator. Here, we only sketch the derivations, whereas the details can be found in the Supplemental Material Sec.~VI.

Our approach is based on identities linking the Fourier transformation of the projectors onto the EPR-like
states (\ref{EPR}) and von Mises states (\ref{vM}) with the ordering of the operators $E$ and $L$:
\begin{eqnarray}
\label{EPRFT}
2\pi\left(\mathcal{F}|n,\alpha\rangle_{sa}\langle n,\alpha|\right)(l,\phi)&=&D_{s}(l,\phi)D_{a}(-l,\phi),\\
\label{vMFT}
\left(\mathcal{F}|n,\alpha\rangle_{s}\langle n,\alpha|\right)(l,\phi)&=&o(l,\phi)D_{s}(l,\phi),
\end{eqnarray}
where
\begin{eqnarray*}
\label{FT}
\left(\mathcal{F}A\right)(l,\phi)=\sum_{n\in\mathbb{Z}}\int_{-\pi}^{\pi}\frac{d\alpha}{2\pi} e^{i(l\alpha-\phi n)}A(n,\alpha),
\end{eqnarray*}
is the Fourier transformation,
\begin{eqnarray}\label{Doperator}
D(l,\phi)=e^{-il\frac{\phi}{2}}E^{-l} e^{-iL\phi}
\end{eqnarray}
is the displacement operator \cite{Rigas_08}, and
\begin{eqnarray}\label{vacoverlap}
o(l,\phi)=e^{il\frac{\phi}{2}}\langle l,\phi|0,0\rangle=\frac{I_l\left[2\kappa \cos\left(\frac{\phi}{2}\right)\right]}{I_0(2\kappa)}.
\end{eqnarray}

The relation (\ref{EPRFT}) follows immediately  from  the orthogonal expansion of the operator $\mathcal{E}^{-l}e^{-i\mathcal{L}\phi}$
in terms of  the  states (\ref{EPR}), whereas the relation (\ref{vMFT}) is obtained by averaging Eq.~(\ref{EPRFT}) over the ancilla von Mises state $|0,0\rangle_{a}$
with  spread parameter $\kappa$. Based on the equality (\ref{vMFT}) we can now construct the  phase-space distributions for angular momentum and angular
variable, in a manner analogous to the quasiprobability $Q$-function \cite{Husimi_40}, Wigner function \cite{Wigner_32} and $P$-function \cite{Glauber_63,Sudarshan_63} of the harmonic oscillator.
In particular, the averaging of the formula (\ref{vMFT}) over the rescaled density operator $\rho/(2\pi)$ yields immediately the relationship between the characteristic function
$C_{Q}(l,\phi)=(\mathcal{F} Q)(l,\phi)$ of the $Q$-function,
\begin{eqnarray}\label{Qfunction}
Q(n,\alpha)=\frac{\langle n,\alpha|\rho|n,\alpha\rangle}{2\pi},
\end{eqnarray}
and the Wigner characteristic function defined as $C_{W}(l,\phi)=\mbox{Tr}\left[\rho D(l,\phi)\right]/2\pi$,
\begin{eqnarray}\label{CQW}
C_{Q}(l,\phi)=o(l,\phi)C_{W}(l,\phi).
\end{eqnarray}

The analogies with the phase-space distributions  of the harmonic oscillator can be taken
further by defining the diagonal representation of a density matrix $\rho$ as $P$-distribution,
analogous to the Glauber-Sudarshan quasiprobability distribution \cite{Sudarshan_63,Glauber_63}.
Recall first  that the displacement operator (\ref{Doperator}) satisfies the following completeness
property \cite{Rigas_08}:
\begin{eqnarray}\label{DcompletenessSM}
\mbox{Tr}\left[D^{\dag}(l,\phi)D(l',\phi')\right]=2\pi\delta_{ll'}\delta_{2\pi}(\phi-\phi'),
\end{eqnarray}
where $\delta_{2\pi}(\phi)$ is the $2\pi$-periodic delta function. Thus, one can express any density matrix as
\begin{eqnarray}\label{rhodecomposition}
\rho=\sum_{l\in\mathbb{Z}}\int_{-\pi}^{\pi}d\phi C_{W}(l,\phi)D^{\dag}(l,\phi).
\end{eqnarray}
Insertion of $[o(l,\phi)]^{-1}o(l,\phi)=1$ into the integrand and application of the unitarity of the Fourier transformation
brings us straightforwardly to the $P$-representation of any density matrix:
\begin{eqnarray}\label{Pfunction}
\rho = \sum_{n\in\mathbb{Z}}\int_{-\pi}^{\pi}d \alpha P(n, \alpha) |n, \alpha\rangle \langle n, \alpha|,
\end{eqnarray}
where we introduced the analogy of the $P$-function as the Fourier transformation
$P(n,\alpha)=(\mathcal{F} C_{P})(n,\alpha)$ of the corresponding characteristic function $C_{P}(l,\phi)$ defined by
\begin{eqnarray}\label{CWP}
C_{W}(l,\phi)&=&o(l,\phi)C_{P}(l,\phi).
\end{eqnarray}
From Eqs.~(\ref{CQW}) and (\ref{CWP}) it is apparent that the ``Bessel'' overlap (\ref{vacoverlap}) plays for the pair of angular momentum and angular
variable exactly the same role of a universal smoothing factor as  the Gaussian overlap $\langle \alpha|0\rangle=\mbox{exp}(-|\alpha|^2/2)$ of the vacuum state $|0\rangle$ and the
coherent state $|\alpha\rangle$ of a harmonic oscillator. The relationship between the respective phase-space distributions is given by the convolution with
the kernel comprised by the Fourier transformation of the overlap (\ref{vacoverlap}). This phase-space  structure and associated  quasi-probability distributions
related to operator ordering constitute the final major result of our Letter.

{\it Quantum communication with von Mises states.}--There were several  experimental attempts to use angular momentum  and angle in a manner analogous to
quadrature operators for the purpose of  quantum information processing  \cite{Leach_10,Erhard_18}.
However the formulation was burdened by periodicity of angular variable or missing an analogue of Bell variables.
The simultaneous measurement of $\mathcal{L}$ and $\mathcal{S}$ with optimal ancillary state  provides full  analogy with the quadrature
heterodyne detection. This allows to translate protocols based on optical quadratures and heterodyne detection into the realm of the $L$ and $S$ variables.
For instance, the coherent state cryptography protocol with heterodyne detection \cite{Weedbrook_04}, which does not require switching of measurement bases,
becomes the analogous no-switching protocol with von Mises states. Another application is obtained if we feed the investigated measurement
with other ancilla states, e.g., comprised by one part of the entangled state (\ref{EPR}) with $N=\Phi=0$. The generalized measurement
then plays the role of the Bell measurement for $L$ and $S$, which can be used for quantum teleportation \cite{Bennett_93,Braunstein_98} of von Mises states.
A generalization of such a protocol allowing teleportation of von Mises states between systems with {\it different} fractional angular momenta is provided in
the Supplemental Material Sec.~IV. Realization of the proposed protocol would extend teleportation of finite superpositions of angular momentum
eigenstates \cite{Wang_15} to the genuine ``continuous-variable'' regime when states spanning entire unbounded state space are teleported.

{\it Optical beams.}--It is a challenging task to implement von Mises states as optical beams  by advanced techniques adopting twisted photons similar to \cite{Molina_Terizza_02,Erhard_18} - either as non-diffracting Bessel or Laguerre-Gauss  modes. Such states would truly play  the role of squeezed-like states carrying information about both complementary  observables of  angular momentum and angular variable. New fascinating progress in   compact generation of  optical angular momentum states \cite{NaturePhys21}   together with optimal usage of information  distributed into continuous and discrete variables    represent a step towards new communication schemes    on robust platform of optical beams.

{\it Phase and  intensity as conjugated variables.}--Although the quantum phase problem has a long history with many pitfalls   \cite{Susskind_64}, the canonical commutation  relation for $\mathfrak{e}$(2) can be modified  to the case of phase and  intensity of the signal field.   Considerations inspired by the analysis of  the phase of complex amplitudes allow to formulate the following two-mode
representation: $L= a^{\dagger}_s a_s - a^{\dagger}_a a_a$ and $E  =  \sqrt{(a_s + a_a^{\dagger})/(a^{\dagger}_s   +  a_a)}$. The phase of the signal field enters through the phase of
the complex amplitude $Y=a_s + a_a^{\dagger},\quad [ Y, Y^{\dagger}]=0.$  However, $L$ and $E$  are  represented here by non-commuting operators and simultaneous detection requires
strategies discussed above.

{\it Conclusion.}--We developed a full quantum description of the canonical pair of angular momentum and angular variable obeying commutation rules associated with the group E(2). A central role is played by the von Mises minimum uncertainty states, allowing the performance of optimal measurement as well as the provision of a phase-space representation of states. Since the optimality is linked to saturable uncertainty relations, our
theory has important metrological consequences and may trigger new experimental techniques oriented to state engineering and detection at quantum limits, fully employing the E(2) symmetry.

\acknowledgments

We thank Jan B\'{\i}lek for help with graphics. J.R.  and  Z.H.  acknowledge the  support  from  the project ApresSF supported by the MEYS, Czech Republic, under the QuantERA programme, which has received funding from the European Union Horizon 2020 research and innovation programme and H2020-FETOPEN-2018-2019-2020-01 StormyTune. The work of H.dG. is supported by NSERC of Canada.

\onecolumngrid
\newpage

\centerline{\bf \Large Supplemental Material}

\vspace{1.0cm}

\twocolumngrid

\section{Modified Bessel function}\label{SM_Modified_Bessel_function}

Here  we review useful formulas to help with some   explicit calculations  involving Bessel functions.
The modified Bessel function of integer order $n$, is defined by the integral formula \cite{Watson_44}
\begin{equation}\label{In}
I_{n}(z)=\frac{1}{2\pi}\int_{-\pi}^{\pi}d\phi e^{z\cos\phi+in\phi}.
\end{equation}
From the definition one can see easily that $I_{n}(z)$ is real for real $z$, and satisfies
\begin{equation}\label{InpropertiesSM}
I_{n}(z)=I_{-n}(z),\quad I_{n}(-z)=(-1)^n I_{n}(z),\quad I_{n}(0)=\delta_{n0}.
\end{equation}
In addition, the modified Bessel functions fulfil the recurrence relations \cite{Watson_44}
\begin{equation}\label{rr1SM}
I_{n-1}(z)-I_{n+1}(z)=\frac{2n}{z}I_{n}(z),
\end{equation}
and
\begin{equation}\label{rr2SM}
I_{n-1}(z)+I_{n+1}(z)=2\frac{d}{dz}I_{n}(z).
\end{equation}
Our calculations with modified Bessel functions are greatly simplified by the addition theorem \cite{Watson_44}
\begin{eqnarray}
\sum_{m\in\mathbb{Z}}(-1)^{m}I_{r+m}(Z)I_{m}(z)e^{im\phi}=e^{ir\psi}I_{r}(\omega),
\end{eqnarray}
where $r\in\mathbb{Z}$ and
\begin{eqnarray}
\omega&=&\sqrt{Z^2+z^2-2Zz\cos\phi}, \nonumber\\
Z-z\cos\phi&=&\omega\cos\psi,\quad z\sin\phi=\omega\sin\psi.
\end{eqnarray}
In particular, the addition formula yields
\begin{eqnarray}\label{Iaddition2}
\sum_{m\in\mathbb{Z}}I_{m}(\kappa)I_{m+r}(\kappa)e^{im\phi}=e^{-ir\frac{\phi}{2}}I_{r}\left[2\kappa\cos\left(\frac{\phi}{2}\right)\right],\nonumber\\
\end{eqnarray}
with the special case
\begin{eqnarray}
\sum_{m\in\mathbb{Z}}I_{m}^{2}(\kappa)=I_{0}(2\kappa).
\end{eqnarray}
The modified Bessel functions can also be obtained from the following generating function \cite{Abramowitz_72}:
\begin{eqnarray}\label{ImgeneratingfSM}
\sum_{m\in\mathbb{Z}}I_{m}(z)e^{im\phi}=e^{z\cos\phi}.
\end{eqnarray}
\\

\section{Properties of von Mises states}\label{SM_Properties_vM_states}

In this section we summarise some useful properties of  the von Mises states, Eq.~(2) of the main text:
\begin{eqnarray}\label{vMSM0}
|n+\delta,\alpha\rangle=\frac{1}{\sqrt{I_0( 2 \kappa) }}\sum_{l\in \mathbb{Z}} e^{i (n -l) \alpha} I_{n-l} (\kappa)|l+\delta\rangle,
\end{eqnarray}
where $\delta\in[0,1)$ and $\kappa\geq0$.

Recall first that  von Mises states (\ref{vMSM0}) are defined as the states saturating
the uncertainty relations (1) of the main text,
\begin{equation}\label{URSM}
\langle(\Delta L)^2\rangle\langle(\Delta S_{\alpha})^2\rangle\ge \frac14 |\langle C_{\alpha}\rangle |^2.
\end{equation}
In the $\phi$-representation von Mises states read
\begin{eqnarray}\label{vMSM}
\psi_{n+\delta,\alpha}(\phi)=\frac{1}{\sqrt{2\pi I_0( 2 \kappa)}}e^{i(n+\delta)\phi+\kappa\cos(\phi-\alpha)},
\end{eqnarray}
where the generating function (\ref{ImgeneratingfSM}) has been used. The states can be seen as a special
type of states introduced previously in \cite{Kastrup2006} given in $\phi$-representation by
\begin{eqnarray}\label{vMSM2}
\tilde{\psi}_{n+\delta,\alpha}^{\sigma}(\phi)=\frac{1}{\sqrt{ 2 \pi I_0(2s)}}e^{i[(n+\delta)(\phi-\alpha)+\sigma\sin(\phi-\alpha)]},\nonumber\\
\end{eqnarray}
where $\sigma=\gamma-is$. The states of (\ref{vMSM2}) can be shown to saturate the uncertainty relations
\begin{equation}\label{URSM}
\langle(\Delta L)^2\rangle\langle(\Delta C_{\alpha})^2\rangle\ge \frac{1}{4}\left(|\langle S_{\alpha}\rangle|^2+|\langle\{\Delta L,\Delta C_{\alpha}\}\rangle|^2\right)
\end{equation}
and their relationship to our states (\ref{vMSM}) is given by
\begin{equation}\label{connection}
\tilde{\psi}_{n+\delta,\alpha-\frac{\pi}{2}}^{-i\kappa}(\phi)=    e^{-i(n+\delta)(\alpha-\frac{\pi}{2})}\psi_{n+\delta,\alpha}(\phi).
\end{equation}
In what follows, it  is advantageous to use the states (\ref{vMSM}) as they represent the ``standard form'' of von Mises states in the $\phi$-representation  with $\gamma=0$ guaranteeing vanishing of the anticommutator mean:
$\langle\{\Delta L,\Delta S_{\alpha}\}\rangle=0$. This form is simpler for calculations yet it captures all essential features of minimum uncertainty states (MUS)
for angular momentum and angular variable.

We start with the overlap $\langle n'+\delta,\alpha' |n+\delta, \alpha\rangle$ of
two von Mises states with the same fractional parts $\delta$. By inserting the resolution of identity $\int_{-\pi}^{\pi}d\phi|\phi\rangle\langle\phi|=\openone$ into the overlap we obtain
\begin{widetext}
\begin{eqnarray}\label{overlap1SM}
\langle n'+\delta, \alpha' |n+\delta, \alpha \rangle&=&\int_{-\pi}^{\pi} d\phi\psi^{\ast}_{n'+\delta,\alpha'}(\phi)\psi_{n+\delta,\alpha}(\phi)\stackrel{1}{=}\frac{1}{I_0(2\kappa)}\int_{-\pi}^{\pi}\frac{d\phi}{2\pi}e^{i(n-n')\phi+2\kappa\cos\left(\frac{\alpha-\alpha'}{2}\right)\cos\left[\phi-\left(\frac{\alpha + \alpha'}{2}\right)\right]}\nonumber\\
&\stackrel{2}{=}&e^{i (n-n')\left(\frac{\alpha + \alpha'}{2}\right)}\frac{I_{n-n'}\left[2\kappa\cos\left(\frac{\alpha - \alpha'}{2}\right)\right]}{I_0(2\kappa)},
\end{eqnarray}
\end{widetext}
where to get equality 1, Eq.~(\ref{vMSM}) and the identity $\cos(\phi-\alpha)+\cos(\phi-\alpha')=2\cos\left[\phi-\left(\frac{\alpha + \alpha'}{2}\right)\right]\cos\left(\frac{\alpha - \alpha'}{2}\right)$ were used,
whereas in equality $2$ we used the definition (\ref{In}). Alternatively, the overlap formula can be derived using the definition (\ref{vMSM0}) and the addition theorem (\ref{Iaddition2}) as
\begin{widetext}
\begin{eqnarray}\label{overlap2SM}
\langle n'+\delta, \alpha'| n+\delta, \alpha  \rangle&=&\frac{e^{i(n \alpha -n'\alpha')}}{I_0( 2 \kappa)}\sum_{l\in\mathbb{Z}} e^{i l (\alpha'-\alpha) }I_{n-l}(\kappa)I_{n'-l}(\kappa)=e^{i (n-n')\left(\frac{\alpha + \alpha'}{2}\right)}\frac{I_{n-n'}\left[2\kappa\cos\left(\frac{\alpha - \alpha'}{2}\right)\right]}{I_0(2\kappa)}.
\end{eqnarray}
\end{widetext}
Interestingly, since $I_{n}(0)=\delta_{n0}$ according to the last of equations (\ref{InpropertiesSM}), von Mises states with $\alpha'=\alpha+(2k+1)\pi$, $k\in\mathbb{Z}$, and $n\ne n'$ are orthogonal.
Thus, contrary to the usual intuition, the over-complete von Mises-state basis contains not only nonorthogonal, but also orthogonal states. A more generic overlap formula for states (\ref{vMSM2})
with generally different fractional parts can be found in \cite{Kastrup2006}.

The overlap formula (\ref{overlap1SM}) together with the addition theorem (\ref{Iaddition2}) allow us to calculate arbitrary moments of von Mises states.
To show this, let us note first how the operators $\mbox{exp}(-iL\phi)$ and $E^{-l}$ act on von Mises states (\ref{vMSM0}),
\begin{eqnarray}\label{ElexpLSM}
e^{-iL\phi}|n+\delta,\alpha\rangle&=&e^{-i(n+\delta)\phi}|n+\delta,\alpha+\phi\rangle,\nonumber\\
E^{-l}|n+\delta,\alpha\rangle&=&|n+l+\delta,\alpha\rangle,
\end{eqnarray}
where in derivation of the second equality the relation $E^{\dag}|n+\delta\rangle=|n+1+\delta\rangle$ has been used. Let us now adopt a conventional definition of the moment generating
function of a quantum state $\rho$ as a mean $G(l,\phi)=\mbox{Tr}[\rho \tilde{D}(l,\phi)]$ of the operator
\begin{eqnarray}\label{DoperatorSM}
\tilde{D}(l,\phi)=E^{-l} e^{-iL\phi}.
\end{eqnarray}
Making use of equations (\ref{ElexpLSM}) and the overlap formula (\ref{overlap1SM}) one can show easily, that the moment generating function for the von Mises state
$|n+\delta,\alpha\rangle$ is given by
\begin{eqnarray}\label{GSM}
G(l,\phi)=e^{il\alpha}e^{-i\left(n+\delta-\frac{l}{2}\right)\phi}\frac{I_{l}\left[2\kappa\cos\left(\frac{\phi}{2}\right)\right]}{I_0(2\kappa)}.
\end{eqnarray}
From here one can then get straightforwardly all moments as derivatives
\begin{eqnarray}\label{jointmomSM}
\left\langle E^{-l}L^{N}\right\rangle=i^N\left.\frac{d^{N}}{d\phi^{N}}G(l,\phi)\right\vert_{\phi=0}.
\end{eqnarray}
For $N=0$ we can combine equations (\ref{GSM}) and (\ref{jointmomSM}) to get
\begin{eqnarray}\label{ElSM}
\langle E^{-l}\rangle=\left.G(l,\phi)\right\vert_{\phi=0}=e^{il\alpha}\frac{I_{l}(2\kappa)}{I_{0}(2\kappa)}.
\end{eqnarray}
Moving to $N>0$, let us now express the $N$-th derivative on the right-hand
side of equation (\ref{jointmomSM}) as $i^{N-1}(d^{N-1}/d\phi^{N-1})i(d/d\phi)G(l,\phi)$, calculate the first derivative
$i(d/d\phi)G(l,\phi)$ with the help of generating function (\ref{GSM}) and use the recurrence relation (\ref{rr2SM}) to express the resulting formula for the first
derivative in terms of $G(l,\phi)$ and $G(l\pm1,\phi)$. This yields the $N$-th derivative of the generating function as a linear combination of
$(N-1)$-st derivatives of the generating functions $G(l,\phi)$ and $G(l\pm1,\phi)$, which in turn leads, when combined with the formula (\ref{jointmomSM}), to
the following recurrence relation for the von Mises states:
\begin{widetext}
\begin{eqnarray}\label{momrrSM}
\left\langle E^{-l}L^{N}\right\rangle&=&\frac{\kappa}{4}\left\{e^{i\alpha}\left\langle E^{-(l-1)}\left[L^{N-1}-\left(L-\openone\right)^{N-1}\right]\right\rangle-e^{-i\alpha}\left\langle E^{-(l+1)}\left[L^{N-1}-\left(L+\openone\right)^{N-1}\right]\right\rangle\right\}\nonumber\\
&&+\left(n+\delta-\frac{l}{2}\right)\left\langle E^{-l}L^{N-1}\right\rangle.
\end{eqnarray}
\end{widetext}
Hence, we can rederive moments of the angular momentum \cite{Kastrup2006}
\begin{eqnarray}\label{LmomentsSM}
\langle L\rangle = n+\delta,\quad \langle L^{2}\rangle= (n+\delta)^{2}+\frac{\kappa}{2}\frac{I_{1}(2\kappa)}{I_{0}(2\kappa)},
\end{eqnarray}
and
\begin{eqnarray}\label{DeltaLSM}
\langle(\Delta L)^2\rangle=\frac{\kappa}{2}\frac{I_{1}(2\kappa)}{I_{0}(2\kappa)},
\end{eqnarray}
or derive new moments, e.g.,
\begin{eqnarray}\label{E2DeltaLSM}
\left\langle (E)^{\pm 2}\Delta L\right\rangle&=&\pm e^{\mp i2\alpha}\frac{I_{2}(2\kappa)}{I_{0}(2\kappa)}
\end{eqnarray}
and
\begin{eqnarray}\label{E2DeltaL2SM}
\left\langle (E)^{\pm 2}(\Delta L)^{2}\right\rangle=\frac{e^{\mp i2\alpha}}{2I_{0}(2\kappa)}\left[I_{2}(2\kappa)+\kappa I_{1}(2\kappa)\right],
\end{eqnarray}
where $(E)^{\pm 2}$ stands for the $(\pm 2)$-nd power of $E$. Later in this Supplemental material we use the latter joint moments
to calculate the joint moment appearing in an alternative approach to simultaneous detection of incompatible observables put forward by She and
Heffner \cite{She_Heffner}.

Before doing this, let us briefly comment on another interesting property of von Mises states, which stems from the
recurrence relation (\ref{momrrSM}). Namely, as $\langle L\rangle = n+\delta$ for von Mises states, the joint moment $\left\langle E^{-l}L^{N}\right\rangle$
can be expressed via the mean $\langle L\rangle$ and joint moments involving at most $(N-1)$-st power of the angular momentum operator. Repeated application
of the recurrence relation (\ref{momrrSM}) on the moments on right-hand side thus allows us to express any joint moment $\left\langle E^{-l}L^{N}\right\rangle$
only in terms of powers of the mean value $\langle L\rangle$ and the moments of powers of the operator $E$. This can be viewed as an analogy of a similar property of
Gaussian quantum states \cite{Weedbrook_12}. These states are fully determined by the first-order and second-order moments of the quadrature operators and
thus any higher-order moment can be expressed only in terms of the first two moments.

\section{Simultaneous detection of angular momentum and angular variable}\label{SM_Simultaneous_detection}

In this section we derive a fundamental lower bound for product of variances of the outcomes of simultaneous measurements of the non-commuting observables
$L_{s}$ and $S_{s}$. This can be done most easily using a joint measurement of commuting observables
$\mathcal{L}=L_{s}+L_{a}$ and $\mathcal{S}=(\mathcal{E}^\dag -\mathcal{E})/2i$ of the signal $s$ and ancilla $a$, where $\mathcal{E}=E_{s}E_{a}^{\dag}$. We seek the measurement which would reach
the minimum of the uncertainty product $\langle(\Delta\mathcal{L})^{2}\rangle\langle(\Delta\mathcal{S})^{2}\rangle$ over the product state $|\varphi\rangle_s|\chi\rangle_a$ with $\Delta\mathcal{S}=\mathcal{S}_{\beta=\mbox{arg}\langle E_{a}\rangle-\mbox{arg}\langle E_{s}\rangle}$ and
$\mathcal{S}_{\beta}=(e^{-i\beta}\mathcal{E}^\dag - e^{i\beta}\mathcal{E})/2i$, acting on both the ancilla and system spaces.
We require the measurement to preserve the signal angular momentum mean, i.e., $\langle \mathcal{L}\rangle=\langle {L}_{s}\rangle$, as well as the angular dependence of relevant
moments of angular variable, $\langle\mathcal{S}^{l}\rangle$, $l=1,2$, on the signal state. This restricts the ancilla state as
\begin{equation}\label{unbiasedness}
\langle  L_a \rangle =0, \quad \arg \langle E_a \rangle = 0, \quad \arg \langle E_a^{2} \rangle = 0.
\end{equation}
Making use of the latter two conditions one can cast the uncertainties  of angular  variables in the form:
\begin{eqnarray}
\langle(\Delta\mathcal{S})^{2}\rangle&=&\frac{1}{2}\left(1-e_{s}e_{a}\cos\psi_{s}\right),\quad \langle(\Delta S_a)^2\rangle=\frac{1}{2}(1 - e_a),\nonumber\\
\langle(\Delta S_s)^2\rangle&=&\frac{1}{2}\left(1-e_{s}\cos\psi_{s}\right),
\end{eqnarray}
where $e_{s,a}=|\langle E_{s,a}^2\rangle|$ and $\psi_{s}=2\arg \langle E_s\rangle-\arg\langle E_s^{2}\rangle$.
Consequently, the measurable uncertainties are simply given as
\begin{eqnarray}
\langle(\Delta\mathcal{L})^2\rangle &=&\langle(\Delta L_s)^2\rangle +\langle(\Delta L_a)^2\rangle, \nonumber\\
\langle(\Delta\mathcal{S})^2\rangle&=&\langle(\Delta S_a)^2\rangle + e_a\langle(\Delta S_s)^2\rangle.
\end{eqnarray}
Hence, the uncertainty product is lower bounded as
\begin{widetext}
\begin{eqnarray}\label{result2SM}
\langle(\Delta\mathcal{L}) ^2\rangle\langle(\Delta\mathcal{S})^2\rangle&=&\left[\langle(\Delta L_s)^2\rangle + \langle(\Delta L_a)^2\rangle\right]\left[\langle(\Delta S_a)^2\rangle
+e_a\langle(\Delta S_s)^2\rangle\right]\nonumber\\
&=&\langle(\Delta L_a)^2\rangle\langle(\Delta S_a)^2\rangle+e_a\langle(\Delta L_s)^2\rangle\langle(\Delta S_s)^2\rangle
+\langle(\Delta L_s)^2\rangle\langle(\Delta S_a)^2\rangle+e_a\langle(\Delta L_a)^2\rangle\langle(\Delta S_s)^2\rangle\nonumber\\
&\stackrel{1}{\geq}&\left[\sqrt{\langle(\Delta L_a)^2\rangle\langle(\Delta S_a)^2\rangle}+\sqrt{e_a\langle(\Delta L_s)^2\rangle\langle(\Delta S_s)^2\rangle}\right]^2\nonumber\\
&\stackrel{2}{\geq}&\frac{1}{4}\left(|\langle E_a\rangle| + |\langle E_s\rangle |\sqrt{|\langle E_{a}^2\rangle|}\right)^2.
\end{eqnarray}
\end{widetext}
Here, inequality $1$ follows from
\begin{eqnarray}
\left[\sqrt{\langle(\Delta L_s)^2\rangle\langle(\Delta S_a)^2\rangle}-\sqrt{e_a\langle(\Delta L_a)^2\rangle\langle(\Delta S_s)^2\rangle}\right]^2\geq0\nonumber\\
\end{eqnarray}
and it is saturated if
\begin{eqnarray}
\label{kappaconditionSM}
\langle(\Delta L_s)^2\rangle\langle(\Delta S_a)^2\rangle = e_a \langle(\Delta L_a)^2\rangle\langle(\Delta S_s)^2\rangle.
\end{eqnarray}
The inequality $2$ is a consequence of the uncertainty relations
\begin{eqnarray}
\langle(\Delta L_{s,a}) ^2\rangle\langle(\Delta S_{s,a})^2\rangle\geq\frac{1}{4}|\langle E_{s,a}\rangle|^2,
\end{eqnarray}
and it is saturated by the von Mises MUS of both the signal and the ancilla. For the condition of Eq.~(\ref{kappaconditionSM}) to hold, we will see that the parameters $\kappa_a$ and $ \kappa_s$  for these states must be related. Recall that for a general von Mises state
$|n+\delta,\alpha\rangle$ one has $\langle L\rangle = n+\delta$ and $\langle E^{l}\rangle=\mbox{exp}(-il\alpha)
I_{l}(2\kappa)/I_{0}(2\kappa)$, Eqs.~(\ref{LmomentsSM}) and (\ref{ElSM}), and the unbiasedness conditions (\ref{unbiasedness}) imply the optimal ancilla state to be the von Mises ``vacuum'' state
$|0,0\rangle_{a}$. If the condition $\langle L_a \rangle   = 0$ is relaxed, the optimal ancilla state
reads $|\delta_{a},0\rangle_{a}$, where $\delta_{a}\in[0,1)$. Similarly, the optimal signal state is also a von Mises state $|n+\delta_{s},\alpha\rangle_{s}$. What is more, substituting the variances for signal and ancilla von Mises states,
\begin{equation}
\langle(\Delta L_{j})^2\rangle=\frac{\kappa_{j}}{2}\frac{I_{1}(2\kappa_{j})}{I_{0}(2\kappa_{j})},
\quad  \langle(\Delta S_{j})^2\rangle=\frac{1}{2\kappa_{j}}\frac{I_{1}(2\kappa_{j})}{I_{0}(2\kappa_{j})},
\end{equation}
$j=s,a $, into the condition of Eq.~(\ref{kappaconditionSM}) one finds that the signal and ancilla spread parameters $\kappa_{s}$ and
$\kappa_{a}$ of optimal states must fulfill the non-trivial condition
\begin{eqnarray}\label{conditionSM}
\kappa_s=\sqrt{\frac{I_{2}(2\kappa_{a})}{I_{0}(2\kappa_{a})}}\kappa_{a}.
\end{eqnarray}
Thus in accordance with our intuition, it is optimal to carry out a von Mises measurement on von Mises states,
but contrary to our intuition, the spread parameter of the ancilla $\kappa_{a}$ and of the measured state
$\kappa_s $ differ.

\section{Quantum teleportation of von Mises states}\label{SM_Quantum_teleportation}

This section deals with the unconditional teleportation of von Mises states. Let us consider two quantum systems $A$ and $B$ with angular momenta $L_{A}$ and $L_{B}$,
and angular variables $E_{A}$ and $E_{B}$, respectively. The simultaneous measurement of the total orbital angular momentum
$\mathcal{L}=L_{A}+L_{B}$ and of the  sine of the angular difference $\mathcal{S}=(\mathcal{E}^\dag -\mathcal{E})/2i=(E_{A}^{\dag}E_{B}-E_{A}E_{B}^{\dag})/2i$
plays in the optimal simultaneous measurement of $L_{A}$ and $S_{A}$ the same role as the EPR operators $x_{A}-x_{B}$ and $p_{A}+p_{B}$ in the optimal simultaneous measurement
of $x_{A}$ and $p_{A}$. Since the latter measurement is nothing but the Bell measurement for continuous-variable systems
\cite{Braunstein_98}, one expects that the former measurement will realize the Bell measurement for orbital angular momentum and
angular variable. In the following we confirm this by showing that the measurement can be used for perfect quantum teleportation \cite{Bennett_93} of unknown von Mises states.

Assume the two systems $A$ and $B$ under consideration carry generally different angular momenta characterised by
fractional parts $\delta_{A}$ and $\delta_{B}$, respectively. Consider further the vectors
\begin{widetext}
\begin{equation}\label{EPRSM}
|N+\Delta_{AB},\Phi\rangle_{AB}=\frac{1}{\sqrt{2\pi}}\sum_{l\in\mathbb{Z}}e^{-i l \Phi}|l+\delta_{A}+N-I_{AB}\rangle _A|-l+\delta_{B}\rangle_B,
\end{equation}
\end{widetext}
where $I_{jk}=[\delta_{j}+\delta_{k}]\in\{0,1\}$ and $\Delta_{jk}=(\delta_{j}+\delta_{k})\,\mbox{mod}\,1$, $\Delta_{jk}\in[0,1)$, $j,k=A,B$, are the
integer part and the fractional part of $\delta_{j}+\delta_{k}$, respectively. The normalisation factor $1/\sqrt{2\pi}$ ensures
that the states are normalized as
\begin{equation}\label{EPRnormSM}
\langle M+\Delta_{AB},\Psi|N+\Delta_{AB},\Phi\rangle=\delta_{MN}\delta_{2\pi}(\Psi-\Phi),
\end{equation}
where
\begin{equation}\label{combSM}
\delta_{2\pi}(\phi)=\frac{1}{2\pi}\sum_{n\in\mathbb{Z}}e^{in\phi}=\sum_{n\in\mathbb{Z}}\delta(\phi-2n\pi)
\end{equation}
is the $2\pi$-periodic delta function (or Dirac comb).
From relations $E|n+\delta\rangle=|n-1+\delta\rangle$ and $E^{\dag}|n+\delta\rangle=|n+1+\delta\rangle$ it further follows
straightforwardly, that
\begin{eqnarray}\label{EPReigenstatesSM}
\mathcal{L}|N+\Delta_{AB},\Phi\rangle_{AB}&=&(N+\Delta_{AB})|N+\Delta_{AB},\Phi\rangle_{AB},\nonumber\\
\mathcal{E}|N+\Delta_{AB},\Phi\rangle_{AB}&=&e^{-i\Phi}|N+\Delta_{AB},\Phi\rangle_{AB},\nonumber\\
\mathcal{E}^{\dag}|N+\Delta_{AB},\Phi\rangle_{AB}&=&e^{i\Phi}|N+\Delta_{AB},\Phi\rangle_{AB},
\end{eqnarray}
where $\mathcal{E}=E_{A}E_{B}^{\dag}$, and thus the vectors of Eq.~(\ref{EPRSM}) are common eigenvectors of $\mathcal{L}$ and $\mathcal{S}$ corresponding to eigenvalues
$(N+\Delta_{AB})$ and $\sin\Phi$, respectively.

Adopting the line of argument of Ref.~\cite{Hofmann_00} we can now design the following teleportation protocol.
The goal of the protocol is to transmit faithfully an unknown von Mises state $|n+\delta_{\rm in},\alpha\rangle_{\rm in}$ of
an input system ``in'' characterized by the fractional part of angular momentum $\delta_{\rm in}$, from a sender Alice to a
receiver Bob. For this purpose, the participants can use the shared ``EPR-like'' state  of Eq.~(\ref{EPRSM})
\begin{eqnarray}\label{EPRSM00}
|\Delta_{AB},0\rangle_{AB}=\frac{1}{\sqrt{2\pi}}\sum_{l\in\mathbb{Z}}|l+\delta_{A}-I_{AB}\rangle _A|-l+\delta_{B}\rangle_B\nonumber\\
\end{eqnarray}
corresponding to eigenvalue $\Delta_{AB}$ of $\mathcal{L}$ and zero eigenvalue of $\mathcal{S}$. First, Alice performs measurement of   EPR-like  states (\ref{EPRSM})
on subsystem ``in'' and her part $A$ of the shared state. Provided that the outcomes of her measurement are $(M,\Psi)$, the global state $|n+\delta_{\rm in},\alpha\rangle_{\rm in}|\Delta_{AB},0\rangle_{AB}$ collapses to the
(unnormalized) state
\begin{widetext}
\begin{eqnarray}\label{collapsed}
_{{\rm in}A}\langle M+\Delta_{{\rm in} A},\Psi|n+\delta_{\rm in},\alpha\rangle_{\rm in}|\Delta_{AB},0\rangle_{AB}&=&\frac{e^{i(I_{AB}-\delta_{B})\Psi}}{2\pi}e^{-i(M+I_{AB}-I_{{\rm in}A})\Psi}E_{B}^{M+I_{AB}-I_{{\rm in}A}}e^{iL_{B}\Psi}|n+\delta_{B},\alpha\rangle_{B}\nonumber\\
&=&\frac{e^{i(I_{AB}+I_{{\rm in}A}-M-2\delta_{B})\frac{\Psi}{2}}}{2\pi}D_{B}^{-1}(M+I_{AB}-I_{{\rm in}A},\Psi)|n+\delta_{B},\alpha\rangle_{B},
\end{eqnarray}
\end{widetext}
where
\begin{eqnarray}\label{DisplacementSM}
D(l,\phi)=e^{-il\frac{\phi}{2}}E^{-l} e^{-iL\phi}
\end{eqnarray}
is the displacement operator \cite{Rigas_08}, and where in the second equality we used the relation
\begin{equation}\label{inverseDSM}
E^{l}e^{iL\phi}=e^{il\phi}e^{iL\phi}E^{l}=e^{il\frac{\phi}{2}}D^{-1}(l,\phi).
\end{equation}
Alice subsequently sends the outcomes of her measurement to Bob via classical channel
and he applies to his part of the shared state the correcting operation $D_{B}(M+I_{AB}-I_{{\rm in}A},\Psi)$.
Up to an irrelevant phase factor and generally different fractional part $\delta_{B}$ from $\delta_{\rm in}$, Bob recreates
a perfect replica $|n+\delta_{B},\alpha\rangle_{B}$ of the original von Mises state on his system and thus he completes the teleportation.

The result above strengthens the attractiveness of a  laboratory implementation of the von Mises measurement. First, the measurement would allow teleportation of von Mises states thereby extending teleportation of finite superpositions of angular momentum eigenstates \cite{Wang_15} to the ``continuous-variable'' regime in which infinite superpositions of angular momentum eigenstates, which span entire infinite-dimensional Hilbert state space, are teleported. In addition, the presented protocol  allows, at least in principle, to teleport
quantum states between systems with generally different fractional angular momenta. It can be expected  that the utility of von Mises measurement will also further carry over to all other translations of quantum information protocols to angular momentum - angle, which utilize Bell measurement,  such as  entanglement swapping \cite{Zukowski_93} or quantum cryptography without measurement switching \cite{Weedbrook_04}.

Note finally, that here we demonstrated perfect teleportation of von Mises states  using the non-normalizable  EPR-like state  of Eq.~(\ref{EPRSM}). Analysis of the realistic protocol with physical approximation of the state (\ref{EPRSM00}), such as, for instance, the entangled state $\sum_{l\in\mathbb{Z}}c_{l,-l}|l\rangle_{A}|-l\rangle_{B}$ generated in the process of spontaneous parametric down-conversion \cite{Mair_01}, is outside the immediate scope of the present work.

\section{She-Heffner approach to simultaneous measurement}\label{SM_She_Heffner_approach}

This section contains analysis of the simultaneous detection of the angular momentum and angular variable based on the statistical perspective introduced
in the seminal paper of She and Heffner \cite{She_Heffner}. Following their argumentation simultaneous detection can be cast as a two-stage process -
state preparation specified by the moments and repeated detection conditioned by the same constraints as in the state preparation step.

The EPR-like states  of Eq.~(\ref{EPRSM}) allow us to bridge the Arthurs-Kelly and She-Heffner approaches. Note first that the states (\ref{EPRSM}) satisfy the completeness condition
\begin{eqnarray}\label{EPRcompletenessSM}
\sum_{N\in\mathbb{Z}}\int_{-\pi}^{\pi}d\Phi|N+\Delta_{sa},\Phi\rangle_{sa}\langle N+\Delta_{sa},\Phi|=\openone_{sa},\nonumber\\
\end{eqnarray}
where we have done the following identification $s\equiv A$ and $a\equiv B$. With the help of the resolution of identity and the eigenvalue equations
(\ref{EPReigenstatesSM}) we can express the product of squares of operators $\Delta\mathcal{L}$ and $\Delta\mathcal{S}$ as
\begin{widetext}
\begin{eqnarray}\label{DLDSSM}
(\Delta\mathcal{L})^2(\Delta\mathcal{S})^2=\sum_{N\in\mathbb{Z}}\int_{-\pi}^{\pi}d\Phi(N+\Delta_{sa}-\langle\mathcal{L}\rangle)^2\sin^{2}\left(\Phi-\beta\right)|N+\Delta_{sa},\Phi\rangle_{sa}\langle N+\Delta_{sa},\Phi|,
\end{eqnarray}
\end{widetext}
where once again  $\beta=\mbox{arg}\langle E_{a}\rangle-\mbox{arg}\langle E_{s}\rangle$. Let us now calculate the partial average of the latter operator over the optimal ancilla state $|\delta_{a},0\rangle_{a}$. Taking into account that for this ancilla $\langle L_{a}\rangle=\delta_{a}$, $\mbox{arg}\langle E_{a}\rangle=0$ and $_{a}\langle\delta_{a},0|N+\Delta_{sa},\Phi\rangle_{sa}=|N-I_{sa}+\delta_{s},\Phi\rangle_{s}/\sqrt{2\pi}$ we get after some algebra the following signal operator, diagonal in the von Mises states $|N+\delta_{s},\Phi\rangle_{s}$:
\begin{widetext}
\begin{eqnarray}\label{DLDSaSM}
\langle(\Delta\mathcal{L})^2(\Delta\mathcal{S})^2\rangle_{a}=\sum_{N\in\mathbb{Z}}\int_{-\pi}^{\pi}\frac{d\Phi}{2\pi}(N+\delta_{s}-\langle{L}_{s}\rangle)^2\sin^{2}\left(\Phi+\mbox{arg}\langle E_{s}\rangle\right)|N+\delta_{s},\Phi\rangle_{s}\langle N+\delta_{s},\Phi|,
\end{eqnarray}
\end{widetext}
where  $\langle X_{sa}\rangle_{a}=\!\,_{a}\langle\delta_{a},0|X_{sa}|\delta_{a},0\rangle_{a}$. Further, by averaging the latter operator over the signal state $\rho_{s}$, we get the analogue of the She-Heffner
integral \cite{She_Heffner} for the angular momentum and angular variable:
\begin{widetext}
\begin{eqnarray}\label{SHSM}
\langle(\Delta\mathcal{L})^2(\Delta\mathcal{S})^2\rangle=\sum_{N\in\mathbb{Z}}\int_{-\pi}^{\pi}\frac{d\Phi}{2\pi}(N+\delta_{s}-\langle{L}_{s}\rangle)^2\sin^{2}\left(\Phi+\mbox{arg}\langle E_{s}\rangle\right)_{s}\!\langle N+\delta_{s},\Phi|\rho_{s}|N+\delta_{s},\Phi\rangle_{s}.
\end{eqnarray}
\end{widetext}
Making use of the expressions for the moments given  in Eq.~(\ref{ElSM}) and Eqs.~(\ref{DeltaLSM})-(\ref{E2DeltaL2SM}), we can finally calculate the She-Heffner moment for von Mises states with spread parameters
$\kappa_{s}$ and $\kappa_{a}$ in the form
\begin{widetext}
\begin{equation}\label{SHSM}
\langle(\Delta\mathcal{L})^2(\Delta\mathcal{S})^2\rangle=\frac{1}{4I_{0}(2\kappa_{s})I_{0}(2\kappa_{a})}\left[\left(\frac{\kappa_{s}}{\kappa_{a}}+\frac{\kappa_{a}}{\kappa_{s}}\right)
I_{1}(2\kappa_{s})I_{1}(2\kappa_{a})+2I_{2}(2\kappa_{s})I_{2}(2\kappa_{a})\right].
\end{equation}
\end{widetext}
In Fig.~2 of the main text we plot the properly normalized moment $\langle(\Delta\mathcal{L})^2(\Delta\mathcal{S})^2\rangle/|\langle E_{s}\rangle|^2|\langle E_{a}\rangle|^2$
versus the spread parameter $\kappa_{s}$ and $\kappa_{a}$ satisfying condition (\ref{conditionSM}). The figure reveals that the correlated uncertainties represented by the latter moment lie below the uncorrelated ones (\ref{result2SM}) for the same von Mises states:
\begin{widetext}
\begin{eqnarray}
\langle(\Delta\mathcal{L}) ^2\rangle\langle(\Delta\mathcal{S})^2\rangle&=&\frac{1}{4}\left(|\langle E_a\rangle|+|\langle E_s\rangle |\sqrt{|\langle E_a^{2}\rangle|}\right)^2=\frac{1}{4}\left[\frac{I_{1}(2\kappa_{a})}{I_{0}(2\kappa_{a})}+\sqrt{\frac{I_{2}(2\kappa_{a})}{I_{0}(2\kappa_{a})}}\frac{I_{1}(2\kappa_{s})}{I_{0}(2\kappa_{s})}\right]^2.
\end{eqnarray}
\end{widetext}

\section{Phase-space representation}\label{SM_Phase_space_representation}

In this section we show that von Mises states allow the development of a phase-space representation for angular momentum and angular variable, which closely resembles
the phase-space representation for quadrature operators based on standard coherent states. For the sake of simplicity, we restrict our attention to integer angular momentum, the generalization to the fractional angular momenta being deferred for further research. The key mathematical tool used for the development of the phase-space methods is the Fourier transformation \cite{Plebanski_00}
\begin{eqnarray}\label{FTSM}
\left(\mathcal{F}A\right)(l,\phi)=\sum_{n\in\mathbb{Z}}\int_{-\pi}^{\pi}\frac{d\alpha}{2\pi} e^{i(l\alpha-\phi n)}A(n,\alpha)
\end{eqnarray}
of an operator (or function) $A(n,\alpha)$. Making use of the filtration property of the $2\pi$-periodic delta function (\ref{combSM}) on the interval of
the length $2\pi$, one can show easily that the Fourier transformation (\ref{FTSM}) fulfils the following analogue of the Parseval formula:
\begin{widetext}
\begin{eqnarray}\label{ParsevalSM}
\sum_{l\in\mathbb{Z}}\int_{-\pi}^{\pi}d\phi(\mathcal{F} A)(l,\phi)(\mathcal{F} B)^{\dag}(l,\phi)=\sum_{n\in\mathbb{Z}}\int_{-\pi}^{\pi}d\alpha A(n,\alpha)B^{\dag}(n,\alpha),
\end{eqnarray}
\end{widetext}
where the symbol $^{\dag}$ stands for the Hermitian conjugate. Analogously, one can show that the
Fourier transformation of a product is a convolution of the Fourier transformations of the factors,
\begin{widetext}
\begin{eqnarray}\label{convolutionSM}
\left[\mathcal{F}(AB)\right](n,\alpha)=\sum_{l\in\mathbb{Z}}\int_{-\pi}^{\pi}\frac{d\phi}{2\pi}(\mathcal{F} A)(n-l,\alpha-\phi)(\mathcal{F} B)(l,\phi)=\left(\mathcal{F} A\right)\ast\left(\mathcal{F}B\right)(n,\alpha).
\end{eqnarray}
\end{widetext}
Finally, for $2\pi$-periodic $A(n,\alpha)$ the Fourier transformation (\ref{FTSM}) is also its own inverse.

The phase-space representation relies on the identity linking the ordering of the operators
$\mathcal{L}$  and $\mathcal{E}$ with the Fourier transformation of the projectors onto common eigenstates of the operators, Eq.~(\ref{EPRSM}) with $\Delta_{sa}=0$:
\begin{widetext}
\begin{eqnarray}\label{EPRFTSM}
2\pi\left(\mathcal{F}|n,\alpha\rangle_{sa}\langle n,\alpha|\right)(l,\phi)&=&\mathcal{E}^{-l}e^{-i\mathcal{L}\phi}=D_{s}(l,\phi)D_{a}(-l,\phi),
\end{eqnarray}
\end{widetext}
where $D_{j}(l,\phi)$ is the displacement operator of the subsystem $j=s,a$. The latter relation follows directly from the application of the operator $\mathcal{E}^{-l}e^{-i\mathcal{L}\phi}$
to the resolution of identity for states (\ref{EPRSM}), Eq.~(\ref{EPRcompletenessSM}) with $\Delta_{sa}=0$. Further, by averaging both sides of the equation (\ref{EPRFTSM})
over the von Mises vacuum state $|0,0\rangle_{a}$ of the ancillary system $a$ with spread parameter $\kappa$, we obtain
\begin{eqnarray}\label{vMFTSM}
\left(\mathcal{F}|n,\alpha\rangle_{s}\langle n,\alpha|\right)(l,\phi)&=&o(l,\phi)D_{s}(l,\phi),
\end{eqnarray}
where
\begin{eqnarray}\label{vacoverlapSM}
o(l,\phi)=e^{il\frac{\phi}{2}}\langle l,\phi|0,0\rangle=\frac{I_l\left[2\kappa \cos\left(\frac{\phi}{2}\right)\right]}{I_0(2\kappa)}.
\end{eqnarray}
Here, to get the left-hand side we used $_{a}\langle 0,0|n,\alpha\rangle_{sa}=|n,\alpha\rangle_{s}/\sqrt{2\pi}$, and to calculate the mean $_{a}\langle 0,0|D_{a}(-l,\phi)|0,0\rangle_{a}$
on the right-hand side we used the Eq.~(\ref{GSM}). The Fourier transformation of the projector onto von Mises state (\ref{vMFTSM}) plays a
central role in our approach to development of the phase-space methods for angular momentum and angular variable. An interesting feature of the formula (\ref{vMFTSM})
is the $c$-number function $o(l,\phi)$, Eq.~(\ref{vacoverlapSM}), in front of the displacement operator $D_{s}(l,\phi)$. Below we show, among other things, that for angular momentum and angular variable
the ``overlap'' (\ref{vacoverlapSM}) plays exactly the same role as plays overlap $\langle \alpha|0\rangle=\mbox{exp}(-|\alpha|^2/2)$ of the vacuum state $|0\rangle$ and the coherent state
$|\alpha\rangle$ of a harmonic oscillator.

The relation (\ref{vMFTSM}) allows us to arrive in an elegant way to analogies of the (Husimi) $Q$-function \cite{Husimi_40}, Wigner function \cite{Wigner_32} and Glauber-Sudarshan $P$-function \cite{Sudarshan_63,Glauber_63} of the standard harmonic oscillator. Namely, let us average the relation (with the index $s$ dropped for simplicity)
over the rescaled density operator $\rho/(2\pi)$, i.e.,
\begin{widetext}
\begin{eqnarray}\label{introQSM}
\mbox{Tr}\left[\frac{\rho}{2\pi}\left(\mathcal{F}|n,\alpha\rangle\langle n,\alpha|\right)(l,\phi)\right]=\left[\mathcal{F}\frac{\langle n,\alpha|\rho|n,\alpha\rangle}{2\pi}\right](l,\phi)=o(l,\phi)\frac{1}{2\pi}\mbox{Tr}\left[\rho D(l,\phi)\right].
\end{eqnarray}
\end{widetext}
In analogy with the phase-space distributions of a harmonic oscillator we now introduce the $Q$-function of a density matrix $\rho$ by a prescription
\begin{eqnarray}\label{QSM}
Q(n,\alpha)=\frac{\langle n,\alpha|\rho|n,\alpha\rangle}{2\pi},
\end{eqnarray}
which is normalized as $\sum_{n\in\mathbb{Z}}\int_{-\pi}^{\pi}d\alpha Q(n,\alpha)=1$. Likewise, we define the Wigner characteristic function as the average of the displacement operator,
\begin{eqnarray}\label{CWSM}
C_{W}(l,\phi)=\frac{1}{2\pi}\mbox{Tr}\left[\rho D(l,\phi)\right].
\end{eqnarray}
As the Fourier transformation of the $Q$--function is just its characteristic function, $(\mathcal{F} Q)(l,\phi)=C_{Q}(l,\phi)$, we get from the formula (\ref{introQSM})
the relationship
\begin{eqnarray}\label{CQWSM}
C_{Q}(l,\phi)&=&o(l,\phi)C_{W}(l,\phi).
\end{eqnarray}
Surprisingly, the analogy with the quadrature phase-space can be developed even further. Recall first that the displacement operator (\ref{DisplacementSM}) exhibits the following completeness property \cite{Rigas_08}:
\begin{eqnarray}\label{DcompletenessSM}
\mbox{Tr}\left[D^{\dag}(l,\phi)D(l',\phi')\right]=2\pi\delta_{ll'}\delta_{2\pi}(\phi-\phi').
\end{eqnarray}
The property (\ref{DcompletenessSM}) enables us to decompose any density matrix $\rho$ as
\begin{eqnarray}\label{rhodecomposition1SM}
\rho=\sum_{l\in\mathbb{Z}}\int_{-\pi}^{\pi}d\phi C_{W}(l,\phi)D^{\dag}(l,\phi).
\end{eqnarray}
Consider now the Hermitian conjugate of the equality (\ref{vMFTSM}) (with the index $s$ again dropped)
\begin{eqnarray}\label{vMFTconjugateSM}
\left(\mathcal{F}|n,\alpha\rangle\langle n,\alpha|\right)^{\dag}(l,\phi)&=&o(l,\phi)D^{\dag}(l,\phi).
\end{eqnarray}
By multiplying both sides with $C_{P}(l,\phi)=[o(l,\phi)]^{-1}C_{W}(l,\phi)$ and performing summation over $l$ and integration over $\phi$, we get
\begin{widetext}
\begin{eqnarray}\label{SM}
\sum_{l\in\mathbb{Z}}\int_{-\pi}^{\pi}d\phi C_{P}(l,\phi)\left(\mathcal{F}|n,\alpha\rangle\langle n,\alpha|\right)^{\dag}(l,\phi)&=&\sum_{l\in\mathbb{Z}}\int_{-\pi}^{\pi}d\phi C_{W}(l,\phi)D^{\dag}(l,\phi)=\rho,
\end{eqnarray}
\end{widetext}
where the rightmost equality is a consequence of Eq.~(\ref{rhodecomposition1SM}). If we now apply to the left-hand side the formula
(\ref{ParsevalSM}), we obtain
\begin{eqnarray}\label{PrepresentationSM}
\rho=\sum_{n\in\mathbb{Z}}\int_{-\pi}^{\pi}d\alpha P(n,\alpha)|n,\alpha\rangle\langle n,\alpha|,
\end{eqnarray}
where we defined the $P$-function by the formula
\begin{eqnarray}\label{PSM}
P(n,\alpha)=(\mathcal{F} C_{P})(n,\alpha).
\end{eqnarray}
Equation (\ref{PrepresentationSM}) reveals that any density matrix can be expressed in diagonal form in von Mises states.
This is a direct analogy of the celebrated Glauber-Sudarshan representation \cite{Sudarshan_63,Glauber_63} for the harmonic oscillator.

Summarizing the results, the characteristic functions of different phase-space distributions are related as
\begin{eqnarray}\label{CQWPSM}
C_{Q}(l,\phi)&=&o(l,\phi)C_{W}(l,\phi)=o^{2}(l,\phi)C_{P}(l,\phi).
\end{eqnarray}
The overlap $o(l,\phi)$, Eq.~(\ref{vacoverlapSM}), plays for the pair of angular momentum and angular
variable the same role of a universal ``smoothing'' factor as plays the overlap $\langle \alpha|0\rangle=\mbox{exp}(-|\alpha|^2/2)$
for the canonically conjugate quadrature operators, where \cite{Perina_91}
\begin{eqnarray}\label{CQWPalphaSM}
C_{Q}(\alpha)=e^{-\frac{|\alpha|^2}{2}}C_{W}(\alpha)=e^{-|\alpha|^2}C_{P}(\alpha).
\end{eqnarray}

Application of the Fourier transformation to equation (\ref{CQWPSM}) and utilization of the formula (\ref{convolutionSM}) yield finally the
following relationship between the adjacent phase-space distributions:
\begin{eqnarray}\label{QWPSM}
Q(n,\alpha)&=&\left[(\mathcal{F}o)\ast W\right](n,\alpha),\nonumber\\
W(n,\alpha)&=&\left[(\mathcal{F}o)\ast P\right](n,\alpha).
\end{eqnarray}
We see that the Fourier transformation of the overlap (\ref{vacoverlapSM}) plays the role of a kernel of the
convolution relating different phase-space distributions. As the $P$-function of the von Mises state $|n,\alpha\rangle$ takes the form
\begin{eqnarray}\label{PvMSM}
P^{|n,\alpha\rangle}(m,\beta)=\frac{1}{2\pi}\delta_{nm}\delta_{2\pi}(\alpha-\beta),
\end{eqnarray}
one finds from the second equality of (\ref{QWPSM}) the kernel to be
\begin{eqnarray}\label{FToSM}
(\mathcal{F}o)(n,\alpha)=2\pi W^{|0,0\rangle}(n,\alpha),
\end{eqnarray}
where $W^{|0,0\rangle}(n,\alpha)$ is the Wigner function of the von Mises state $|0,0\rangle$. The
Wigner function is given by a sum of two terms both involving third Jacobi theta function
\cite{Rigas_08} and we can combine it with the formulas (\ref{QWPSM}) and (\ref{FToSM})
to calculate phase-space distributions for other basic states of the investigated system.
This programme as well as further development of the phase-space methods introduced here
is beyond the scope of the present manuscript and will be addressed elsewhere.




\end{document}